\documentclass[sigconf,screen,nonacm]{acmart}

\usepackage[font=small,labelfont=bf]{caption}
\usepackage{subcaption}
\usepackage{prooftree}
\usepackage{soul}
\usepackage{tikz}
\usepackage{siunitx}
\usepackage{color}
\usepackage{listings}
\usepackage{pifont}
\usepackage{multirow}
\usepackage{amsmath}
\usepackage{amsfonts}
\usepackage{amssymb}
\usepackage{graphicx}
\usepackage{paralist}
\usepackage{wrapfig}
\usepackage{balance}
\usepackage{dsfont}
\usepackage{enumitem}
\usepackage{mathtools}
\usepackage{listings}
\usepackage{algorithm}
\usepackage{algpseudocode}
\usepackage{pifont}
\usepackage{stmaryrd}
\usepackage{blindtext}
\usepackage{enumitem}
\usepackage{comment}
\usepackage{url}
\usepackage{etoolbox}
\usepackage{pbox}
\usepackage{csquotes}
\usepackage{ctable} 
\usepackage{textcase}
\usepackage{wasysym}
\usepackage[framemethod=tikz]{mdframed}
\newcommand{\encs}{\blacktriangleright}

\newcommand\myhrulefill[1]{\leavevmode\leaders\hrule height#1\hfill\kern0pt}

\renewcommand{\paragraph}[1]{\vspace{.03in}\noindent\textbf{{#1}~~~}}

\newtoggle{long}

\usepackage{enumitem}

\makeatletter
\lst@InstallKeywords k{attributes}{attributestyle}\slshape{attributestyle}{}ld
\makeatother

\usepackage[scaled=0.80]{DejaVuSansMono}
\lstdefinestyle{customc}{
  belowcaptionskip=1\baselineskip,
  breaklines=true,
  xleftmargin=\parindent,
  language=C,
  showstringspaces=false,
  basicstyle=\ttfamily,
  keywordstyle=\bfseries\ttfamily\color{keywordcolor},
  commentstyle=\itshape\color{darkgray},
  identifierstyle=\ttfamily\color{black},
  stringstyle=\itshape\color{ACMDarkBlue},
  keywords={ public, void, new, action if, else, action, data, category, Intent, Uri},
attributestyle = \bfseries\ttfamily\color{attributecolor}
}

\setlist[itemize]{
   topsep=.5ex,
   itemsep=0ex,
   leftmargin=1em,
    itemindent=0em
}

\setlist[enumerate]{
  topsep=0.5ex,
  itemsep=0pt
}
\setlist[description]{
   labelindent=.3cm,
  style=unboxed,
  leftmargin=.3cm,
  topsep=2pt,
  itemsep=0ex
}

\newcommand{\aws}[1]{{\color{ACMBlue}{\noindent\textsf{\textbf{Aws}---#1}}}}
\newcommand{\jinman}[1]{{\color{ACMPurple}{\noindent\textsf{\textbf{Jinman}---#1}}}}
\newcommand{\vaibhav}[1]{{\color{ACMRed}{\noindent\textsf{\textbf{{Vaibhav}}---#1}}}}
\newcommand{\somesh}[1]{{\color{ACMGreen}{\noindent\textsf{\textbf{{Somesh}}---#1}}}}
\newcommand{\note}[1]{
{\color{red}{\noindent{ \textbf{Notes:} #1}}}
}

\newcommand{\todo}[1]{
{\color{red}{\noindent{ \textbf{TODO:} #1}}}
}

\renewcommand{\aws}[1]{}
\renewcommand{\jinman}[1]{}
\renewcommand{\vaibhav}[1]{}
\renewcommand{\somesh}[1]{}
\renewcommand{\note}[1]{}
\renewcommand{\todo}[1]{}

\everypar{\looseness=-1}

\definecolor{lightgray}{gray}{0.9}
\definecolor{midgray}{gray}{0.65}
\definecolor{darkgray}{gray}{0.4}

\newcommand{\abr}[1]{\textsc{\MakeLowercase{#1}}}

\newcommand{\abrs}[1]{\abr{#1}{\footnotesize{s}}\xspace}

\renewcommand{\vec}[1]{\boldsymbol{#1}}

\newcommand{\act}{\emph{act}\xspace}
\newcommand{\cats}{\emph{cats}\xspace}
\newcommand{\acts}{\emph{acts}\xspace}
\newcommand{\strings}{{\Sigma^*}}
\newcommand{\abs}[1]{#1^\#}

\newcommand{\rone}{(\emph{i})~}
\newcommand{\rtwo}{(\emph{ii})~}

\definecolor{keywordcolor}{gray}{0.0}
\definecolor{attributecolor}{gray}{0.0}
\definecolor{wildcolor}{gray}{0.8}

\usepackage{relsize}

\newcommand{\sgd}{\abr{SGD}\xspace}
\newcommand\earr\hookrightarrow

\newcommand\icc{\abr{icc}\xspace}
\newcommand\icthree{{\abr{ic}\smaller{3}}\xspace}
\newcommand\linn{\abr{linn}\xspace}
\newcommand\linns{\abrs{linn}}
\newcommand\tde{\abr{tde}\xspace}
\newcommand\tdes{\abrs{tde}}
\newcommand\mlp{\abr{mlp}\xspace}

\newcommand\cnn{\abr{cnn}\xspace}
\newcommand\cnns{\abrs{cnn}}

\newcommand\lstm{\abr{lstm}\xspace}
\newcommand\lstms{\abrs{lstm}}

\newcommand\rnn{\abr{rnn}\xspace}
\newcommand\rnns{\abrs{rnn}}

\newcommand\tlstm{Tree-\abr{lstm}\xspace}
\newcommand\tlstms{Tree-\abrs{lstm}}

\newcommand{\R}{\mathbb{R}}

\DeclareMathOperator*{\argmin}{argmin}

\def\1{\mathbbm{1}}

\newcommand{\trianglepara}[1]{\langle{#1}\rangle}
\def\tpara{\trianglepara}

\newcommand{\code}[1]{{\ttfamily{#1}}}
\usepackage{seqsplit}
\newcommand{\longcode}[1]{{\ttfamily\seqsplit{#1}}}

\newcommand{\intents}{I}
\newcommand{\filters}{F}

\newcommand{\intent}{\iota}
\newcommand{\filter}{f}

\newcommand{\pr}{p}
\newcommand{\train}{D}

\newcommand{\echar}{\textsc{e-cat}}
\newcommand{\ereal}{\textsc{e-real}}
\newcommand{\elist}{\textsc{e-list}}
\newcommand{\eset}{\textsc{e-set}}
\newcommand{\eprod}{\textsc{e-prod}}
\newcommand{\esum}{\textsc{e-sum}}

\newcommand{\map}{\emph{map}}
\newcommand{\listt}{\mathrm{L}}
\newcommand{\sett}{\mathrm{S}}
\newcommand{\aggr}{\mathit{aggr}}
\newcommand{\flatten}{\mathit{flat}}
\newcommand{\combine}{\mathit{comb}}

\newcommand{\encchar}{\mathit{enumEnc}}

\def\sigmoid{{\sigma}}
\DeclareMathOperator{\relu}{relu}

\toggletrue{long}
\setcopyright{acmcopyright}
\acmPrice{15.00}
\acmDOI{10.1145/3236024.3236066}
\acmYear{2018}
\copyrightyear{2018}
\acmISBN{978-1-4503-5573-5/18/11}
\acmConference[ESEC/FSE '18]{Proceedings of the 26th ACM Joint European Software Engineering Conference and Symposium on the Foundations of Software Engineering}{November 4--9, 2018}{Lake Buena Vista, FL, USA}
\acmBooktitle{Proceedings of the 26th ACM Joint European Software Engineering Conference and Symposium on the Foundations of Software Engineering (ESEC/FSE '18), November 4--9, 2018, Lake Buena Vista, FL, USA}


\begin{document}

\title[]{Neural-Augmented Static Analysis of Android Communication}         


\author{}




\author{\vspace{-8pt}
Jinman Zhao$^*$ \quad
Aws Albarghouthi$^*$ \quad
Vaibhav Rastogi$^*$ \quad
Somesh Jha$^*$ \quad
Damien Octeau$^\dagger$}
\affiliation{\vspace{-10pt}
$^*$University of Wisconsin-Madison \qquad
$^\dagger$Google \\
jz@cs.wisc.edu \quad
aws@cs.wisc.edu \quad
vrastogi@wisc.edu \quad
jha@cs.wisc.edu \quad
docteau@google.com}

\renewcommand{\authors}{Jinman Zhao, Aws Albarghouthi, Vaibhav Rastogi, Somesh
Jha, and Damien Octeau}
\renewcommand{\shortauthors}{J.~Zhao, A.~Albarghouthi, V.~Rastogi, S.~Jha, and
D.~Octeau}


\begin{abstract}


We address the problem of discovering communication links between
applications in the popular Android mobile operating system, an
important problem for security and privacy in Android.  Any scalable
static analysis in this complex setting is bound to produce an
excessive amount of false-positives, rendering it
impractical.  To improve precision, we propose to augment static
analysis with a trained neural-network model that estimates the
probability that a communication link truly exists.  We describe a
neural-network architecture that encodes abstractions of communicating
objects in two applications and estimates the probability with which a
link indeed exists.  At the heart of our architecture are
\emph{type-directed encoders} (\tde), a general framework for
elegantly constructing encoders of a compound data type by recursively
composing encoders for its constituent types.
We evaluate our approach on a large corpus of Android applications,
and demonstrate that it achieves very high accuracy. Further, we
conduct thorough \emph{interpretability studies} to understand the
internals of the learned neural networks.
\end{abstract}




\maketitle

\section{Introduction}\label{sec:introduction}
In Android, the popular mobile operating system, applications can
communicate with each other using an Android-specific message-passing
system called \emph{Inter-Component Communication} (\icc).  Misuse and
abuse of \icc may lead to several serious security vulnerabilities,
including theft of personal data as well as privilege-escalation
attacks~\cite{eomc11,cfgw11,zj13,gzwj12}.  Indeed, researchers have
discovered various such instances~\cite{cfgw11}---a bus application that
broadcasts \abr{GPS} location to all other applications, an
\abr{SMS}-spying application that is disguised as a tip calculator,
amongst others.
Thus, it is important to detect inter-component communication:
\emph{Can component $c$ communicate with component $d$?} (We say that
$c$ and $d$ have a \emph{link}.)

With the massive size of Android's application market and the
complexity of the Android ecosystem and its applications, a sound and
scalable static analysis for \emph{link inference} is bound to produce
an overwhelmingly large number of false-positive links.
Realistically, however, a security engineer needs to manually
investigate malicious links, and so we need to carefully prioritize
their attention.

To address false positives in this setting, recently \citet{octeau16}
presented \abr{PRIMO}, a hand-crafted probabilistic model that assigns
probabilities to \icc links inferred by static analysis, thus ordering
 links by how likely they are to be true positives.
We see multiple interrelated problems with \abr{PRIMO}'s methodology: First, the model
is \emph{manually} crafted; constructing the model is a laborious,
error-prone process requiring deep expert domain knowledge in the
intricacies of Android applications, the \icc system, and the static
analysis at hand.
Second, the model is specialized; hence, changes in the
Android programming framework---which is constantly evolving---or the
static analysis may render the model obsolete, requiring a new
expert-crafted model.


\begin{figure}[t]
  \includegraphics[scale=1.25]{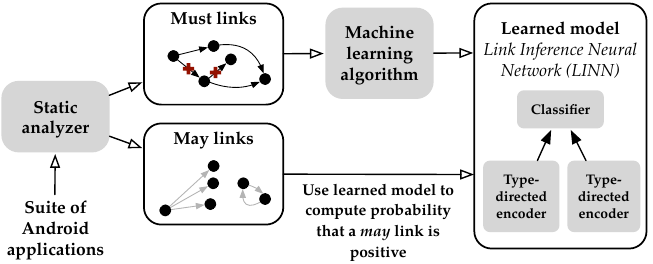}
  \caption{Overview of our approach}
  \vspace{-.1in}
  \label{fig:overview}
\end{figure}

\paragraph{Model-Augmented Link Inference}
Our high-level goal is:
\begin{displayquote}
  \emph{To \textbf{automatically construct} a probabilistic model
  that augments static analysis, \textbf{using minimal domain knowledge}.}
\end{displayquote}

To achieve this, we view the link-inference problem through the lens
of machine learning. We make the observation that we can categorize
results of a static analysis into \emph{must} and \emph{may}
categories: links for which we are sure whether they exist or not
(must links), and others for which we are unsure (may links). We then
utilize must links to train a machine learning classifier, which we
then apply to approximate the likelihood of a may link being a true
positive.  Figure~\ref{fig:overview} presents an overview of our
proposed approach.

\paragraph{Link-Inference Neural Networks}
To enable machine learning in our setting, we propose a custom
neural-network architecture targeting the link-inference problem,
which we call \emph{link-inference neural network} (\linn).  The
advantages of using a neural network for this setting---and generally
in machine learning---
is to automatically extract useful features of link artifacts
(specifically, the addressing mechanism in \icc).  We can train the
network to \emph{encode} the artifacts into real-valued vectors that
capture relevant features.  This relieves us from the arduous and
brittle task of manual feature extraction, which requires \rone expert
domain knowledge and \rtwo maintenance in case of changes in Android
\icc or the used static analysis.

The key novelty of \linns is what we call a \emph{type-directed
  encoder} (\tde): \emph{a generic framework for constructing neural
  networks that encode elements of some type $\tau$ into real-valued
  vectors.}  We use \tdes to encode abstractions of link artifacts
produced by static analysis, which are elements of some data type
representing an \emph{abstract domain}.  \tdes exploit the inherent
recursive structure of a data type: \emph{We can construct an encoder
  for values of a compound data type by recursively composing encoders
  for its constituent types.}  For instance, if we want to encode
elements of the type $\texttt{int}\times\texttt{string}$, then we
compose an encoder of \texttt{int} and an encoder of \texttt{string}.

We demonstrate how to construct encoders for a range of types,
including lists, sets, product types, and strings, resulting in a
generic approach.  Our \tde framework is parameterized by
differentiable functions, which can be instantiated using different
neural-network architectures.  Depending on how we instantiate our
\tde framework, we can arrive at different encoders.  At one extreme,
we can use \tdes to simply treat a value of some data type as its
serialized value, and use a \emph{convolutional} (\cnn) or
\emph{recurrent neural network} (\rnn) to encode it; at the other
extreme, we show how \tdes can be instantiated as a
\tlstm~\cite{tai2015improved}, an advanced architecture that maintains
the tree-like structure of a value of a data type.  This allows us to
systematically experiment with various encodings of abstract states
computed by static analysis.



\paragraph{Contributions}
We make the following contributions:
\begin{itemize}
  \item
  \textbf{Model-Augmented Link Inference (\S\ref{sec:problem}):} We
  formalize and investigate the problem of augmenting a static
  analysis for Android \icc with an automatically learned model that
  assigns probabilities to inferred links.

  \item
\textbf{Link-Inference NN Framework (\S\ref{sec:linn}):} We present a
custom neural-network architecture, \emph{link-inference neural
  network} (\linn), for augmenting a link inference static analysis .
At the heart of \linns are \emph{type-directed encoders} (\tde), a
framework that allows us to construct encoder neural networks for a
compound data type by recursively composing encoders of its component
types.

  \item \textbf{Instantiations \& Empirical Evaluation
    (\S\ref{sec:evaluation}):} We implement our approach using
    TensorFlow~\cite{abadi2016tensorflow} and \icthree~\cite{oljp16}
    and present a thorough evaluation on a large corpus of
    \num[group-separator={,}]{10500} Android applications from the
    Google Play Store.  We present multiple instantiations of our
    \tdes, ranging in complexity and architecture.  Our results
    demonstrate very high classification accuracy.
    Significantly, our automated technique outperforms
    \abr{PRIMO}, which relied on $6$ to $9$ months of manual model engineering.

  \item \textbf{Interpretability Study (\S\ref{sec:explainability}):}
    To address the problem of opacity of deep learning, we conduct a
    detailed \emph{interpretability investigation} to explain the
    behavior of our models.  Our results provide strong evidence that
    our models learn to mimic the link-resolution logic employed by
    the Android operating system.
\end{itemize}

\section{Android ICC: Overview \& Definitions}\label{sec:overview}
\newcommand\must{\mathit{Must}}
\newcommand\may{\mathit{May}}

We now \rone provide  relevant
background about Android \icc,
and
\rtwo formally define \emph{intents}, \emph{filters},
and their static abstractions.

\subsection{ICC Overview and Examples}

Android applications are conceptually collections of \emph{components}, with
each component designed to perform a specific task, such as display a
particular screen to the user or play music files.
Applications can leverage the functionality of other applications
through the use of a sophisticated message-passing system, generally referred to
as \emph{Inter-Component Communication} (\icc). To illustrate, say an application
developer  desires the ability to dial a phone number.  Rather than code that
functionality from scratch, the developer can instead send a message requesting
that some other application handle the dialing process.
Further, such messages are often
generic, i.e., not targeted at a specific application.
The same communication mechanism is also used to send messages within an
application. Consequently,
\emph{any inter-component or inter-application program analysis must first
begin by computing the \abr{ICC} links.}

\begin{figure*}
  \centering
  \begin{subfigure}[b]{.5\textwidth}
  \includegraphics[width=.78\linewidth]{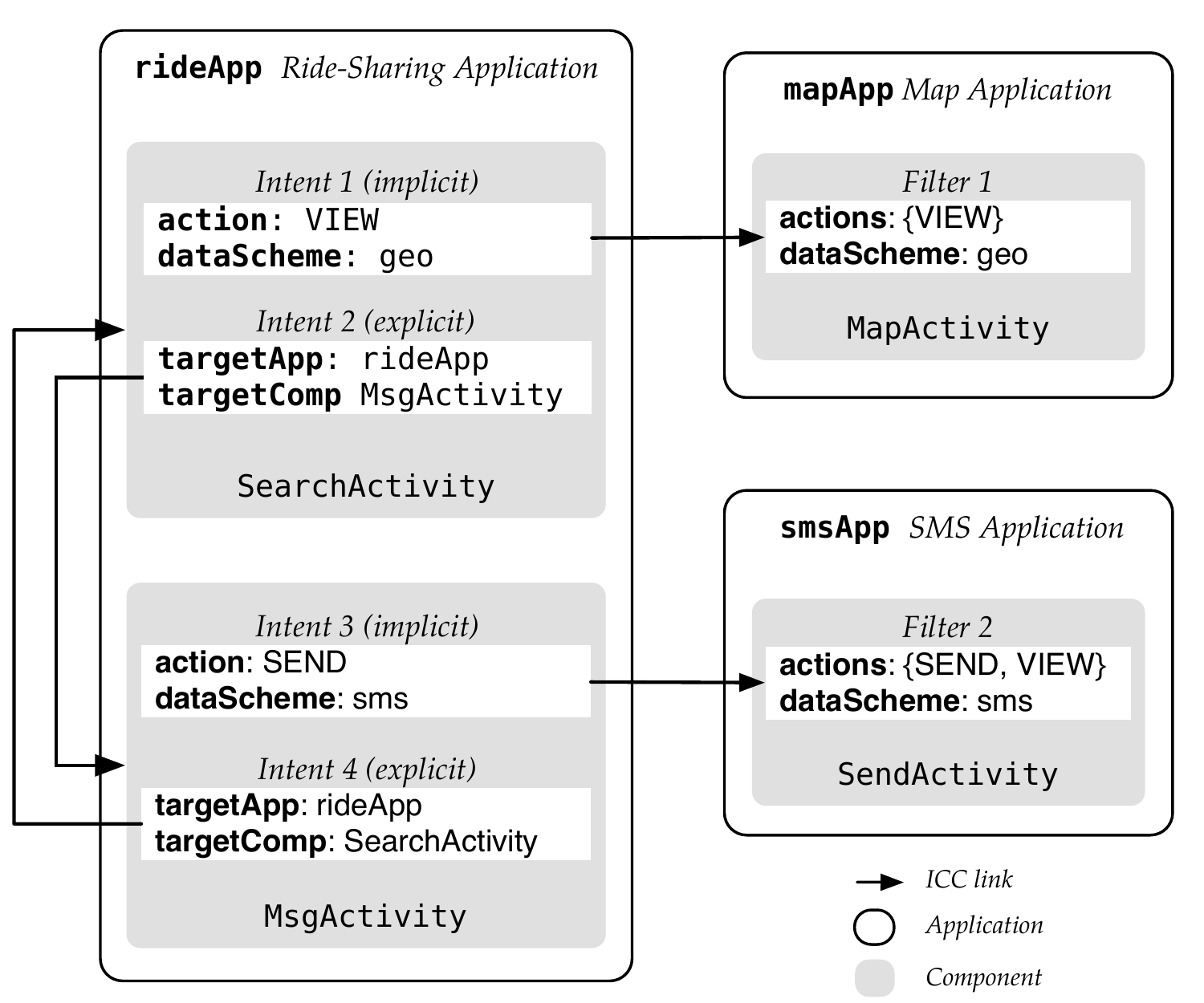}
  \caption{\icc example with three applications}
  \label{fig:icc}
\end{subfigure}
\begin{subfigure}[b]{.4\textwidth}
\small
\begin{lstlisting}
public void sendImplicitIntent() {
    Intent intent = new Intent();
    intent.setAction("SEND");
    msg = ... // contains phone # and msg
    intent.setData(msg);
    startActivity(intent);}
\end{lstlisting}
  \emph{Code constructing and starting implicit intent}

\begin{lstlisting}
<intent-filter>
  <action android:name="SEND"/>
  <action android:name="VIEW"/>
  <data android:scheme="sms"/>
  <category android:name="DEFAULT"/>
</intent-filter>
\end{lstlisting}
  \emph{Intent filter for a \abr{SMS} component}
\caption{Intent for sending an \abr{SMS} and associated filter}
\label{fig:intent-example}
\end{subfigure}
\vspace{-.15in}
\caption{ICC Example}
\end{figure*}

\paragraph{Intents}
Figure~\ref{fig:icc} illustrates a simplified \abr{ICC} example
with a ride-sharing application (\texttt{rideApp}) that uses
functionality of a map application (\texttt{mapApp})
and an \abr{SMS}-messaging application (\texttt{smsApp}).
Each application comprises components (gray boxes),
and arrows in the figure represent potential links between components.
\abr{ICC} is
primarily accomplished using {\em intents}.
Intents are messages sent between Android components.
An intent can be either \emph{explicit} or \emph{implicit}.
The former specifies the target application and component;
the latter merely
specifies the functionality it requires from its target.

Consider, for instance, the implicit intent 3 in Figure~\ref{fig:icc};
it requests the \emph{action} \texttt{SEND} (send an \abr{SMS}),
which sends \texttt{sms} \emph{data}.
By issuing this intent (Figure~\ref{fig:intent-example} top), \texttt{rideApp}
is able to send a message without having to worry about
how this action is performed and by whom.

\paragraph{Intent Filters}
Components that wish to receive implicit intents have to declare {\em intent
filters} (\emph{filters} for short), which describe the attributes of the intents that they are willing to
receive (i.e., subscribe to intents of those types).
For instance, filter 2 in Figure~\ref{fig:icc}
specifies that \texttt{smsApp} can handle \texttt{SEND}
and \texttt{VIEW}
actions (Figure~\ref{fig:intent-example} (bottom) shows the code
declaring this filter).
Therefore, when the ride-sharing application issues an intent
with a \texttt{SEND} action, Android's \emph{intent-resolution} process matches
it with \texttt{smsApp}, which offers that functionality.
Security and privacy issues arise, for instance, when malicious applications are installed
that intercept \abr{SMS} messages by declaring that they also handle
\texttt{SEND} actions~\cite{cfgw11,yxa+15,eyr15}.

%

\subsection{Intents, Filters, and their Abstraction}
We now formalize intents and filters.
We use $\strings$ to denote the set of all strings,
and $\omega$ to denote an undefined value (\texttt{null}).

\begin{definition}[Intents]\label{def:intent}
An \emph{intent} $\intent$ is a pair $(\act,\cats)$, where:
\begin{itemize}
  \item $\act \in \strings \cup \{\omega\}$ is a string defining the \emph{action}
  of $\intent$, e.g., the string \texttt{"SEND"} in Figure~\ref{fig:intent-example} (top), or the value $\omega$.
  \item $\cats \in 2^\strings$ is a set of strings defining \emph{categories},
  which provide further information of how to run the intent.
  For instance,
  \texttt{"APP\_BROWSER"} means that the intent should be able
  to browse the web;
  If an intent is created without providing categories,
  \cats is instantiated into the singleton set $\{\texttt{"DEFAULT"}\}$.
\end{itemize}
Practically, intents also contain the issuing component's name
as well as data like a phone number, image, or email.
We elide these other fields because, based on our dataset, they provided little
information: less than 10\% of intents/filters have fields other than actions and
categories and even when they have them, the values have few distinct
possibilities.
It is, however, conceivable that as we investigate more applications, other 
fields will become important.
\end{definition}

\begin{definition}[Filters]
A \emph{filter} $\filter$ is a tuple $(\acts,\cats)$,
where:
\begin{itemize}
  \item $\acts \in 2^\strings$ is a set of strings defining
   actions a filter can perform.
  \item $\cats \in 2^\strings$ is a set of category strings (see Definition~\ref{def:intent}).
\end{itemize}
Like intents, filters also contain more attributes,
but, for our purposes, actions and categories
are most relevant.
\end{definition}

\paragraph{Semantics}
We will use $\intents$ and $\filters$ to denote
the sets of all possible intents and filters.
We characterize an Android application $A$ by
the (potentially infinite) set of intents $\intents_A \subseteq \intents$
and filters $\filters_A \subseteq \filters$ it can create---i.e.,
the \emph{collecting semantics} of the application.

%

\newcommand{\link}{\emph{match}}
\newcommand{\alink}{\abs{\link}}
\newcommand{\qlink}{\emph{qmatch}}

Let $Y = \{0,1\}$, indicating
whether a link exists (1) or not (0).
We  use the \emph{matching} function $\link: \intents \times \filters \rightarrow Y$
to denote an oracle that, given an intent $\intent$
and filter $\filter$, determines whether $(\intent,\filter)$
can ever match at runtime---i.e., there is a link between them---following the Android intent-resolution
process.
We refer to~\citet{octeau16} for a formal definition
of $\link$.

\begin{example}
Consider the intent and filter described in Figure~\ref{fig:intent-example}.
The intent carries a \texttt{SEND} action, which matches one of the actions
specified in the filter.
The \texttt{msg} in the intent
it has an \texttt{sms} data scheme so that it matches the filter data
scheme.
As for the category, the Android \abr{API} \texttt{startActivity} initializes the categories of  the intent to the set $\texttt{{"DEFAULT"}}$. This then matches the filter's \texttt{"DEFAULT"}
category. Thus, the intent and filter of Figure~\ref{fig:intent-example} match
successfully.
\end{example}

\paragraph{Static Analysis}
We assume that we have a domain of  \emph{abstract intents}
and \emph{abstract filters}, denoted $\abs{\intents}$
and $\abs{\filters}$, respectively.
Semantically, each abstract intent $\abs{\intent} \in \abs{\intents}$
denotes a (potentially infinite) set of intents in $\intents$,
and the same analogously holds for abstract filters.
We assume that we have a static analysis that,
given an application $A$, returns a set
of abstract intents and filters, $\abs{\intents}_A$
and $\abs{\filters}_A$.
These overapproximate the set of possible intents and filters.
Formally: for every $\intent \in \intents_A$, there exists
  an abstract intent $\abs{\intent} \in \abs{\intents}_A$
  such that $\intent \in \abs{\intent}$;
  and analogously for filters.

Provided with an abstract intent and filter---say, from two different apps---%
the static analysis employs an \emph{abstract matching} function $$\alink: \abs{\intents} \times \abs{\filters} \rightarrow \{0,1,\top\},$$
which determines that
   there \emph{must} be link between them (value 1),
   there \emph{must} be no link (value 0),
   or \emph{maybe} there is a link (value $\top$).
The  more imprecise the abstractions
$\abs{\intent},\abs{\filter}$ are, the more
likely that $\alink$ fails
to provide a definitive \emph{must} answer.

We employ the \icthree~\cite{oljp16} tool
for computing abstract intents and filters,
and the \abr{PRIMO} tool~\cite{octeau16} for abstract matching,
which reports \emph{must} or \emph{may} results.
An abstract intent $\abs{\intent}$ computed by \icthree
has the same representation as an intent:
it is a tuple $(\act, \cats)$, except that
strings in $\act$ and $\cats$ are interpreted as regular
expressions.
Therefore, $\abs{\intent}$ represents a \emph{set of intents},
one for each combination of strings
that matches the regular expressions.
The same holds for filters.
For example, an abstract intent can be
the tuple $(\texttt{"(.*)SEND"}, \{\texttt{"DEFAULT"}\})$.
This represents an infinite set of intents where
the action string has the suffix \texttt{"SEND"}.

\section{Model-Augmented Link Inference}\label{sec:problem}

We now view link
inference as a classification problem.

\paragraph{A Probabilistic View of Link Inference}
%
%
%
Recall that $\abs{\intents}$ is the set of abstract intents,
$\abs{\filters}$ is the set of abstract filters, and $Y=\{0,1\}$
indicates whether a link exists.  We can construct a classifier (a
function)
$h: \abs{\intents} \times \abs{\filters} \rightarrow [0,1]$,
which, given an abstract intent $\abs{\intent}$ and
filter $\abs{\filter}$,
$h(\abs{\intent},\abs{\filter})$ indicates the probability that a link
exists (the probability that a link does not exist is
$1-h(\abs{\intent},\abs{\filter})$).  In other words,
$h(\abs{\intent},\abs{\filter})$ is used to estimate the conditional
probability that $y=1$ (link exists) given that the intent is
$\abs{\intent}$ and the filter is $\abs{\filter}$---that is, $h$
estimates the conditional probability $\pr(y \mid \abs{\intent},
\abs{\filter})$.

There are many classifiers used in practice
(e.g. logistic regression).  We focus on neural
networks, as we can use them to automatically address the non-trivial task of encoding intents and filters as vectors of real values---a
process known as \emph{feature extraction}.

\paragraph{Training via Static Analysis}
The two important questions that arise in this context are: \rone how
do we obtain training data, and \rtwo
how do we represent intents and filters in our classifier?

\begin{itemize}
\item \textbf{Training data:}  We make the observation that the
results of a sound static analysis can infer definite
labels---\emph{must} information---%
of the intents and filters it is provided.  Thus, we can \rone randomly sample
 applications from the Android market, and \rtwo use static
analysis to construct a training set $D \subseteq (\abs{\intents} \times \abs{\filters}) \times Y$:
$$\train =
\{\tpara{(\abs{\intent}_1,\abs{\filter}_1),y_1},\ldots,\tpara{(\abs{\intent}_n,\abs{\filter}_n),y_n}\}$$
comprised of intents and filters for which we know the label $y$ with
absolute certainty because they are in the must category for the
static analysis.  \emph{To summarize, we use static analysis as the
  oracle that labels the data for us.}

\item \textbf{Representation:} For a typical machine-learning task,
  the input to the classifier is a vector of real values (i.e.  the
  set of \emph{features}).  Our abstract intents, for example, are
  elements of type $(\Sigma^* \cup \{\omega\}) \times 2^{\Sigma^*}$.
  So, the key question is \emph{how can we transform elements of such
    complex type into a vector of real values?}  In
  \S\ref{sec:linn}, we present a general framework for taking
  some compound data type and training a neural network to encode it
  as a real-valued vector.  We call the technique \emph{type-directed
    encoding}.

\end{itemize}

\paragraph{Augmenting Static Analysis}
Once we have trained a model $h : \abs{\intents} \times \abs{\filters}
\rightarrow [0,1]$, we can compose it with
the results of static analysis to
construct a quantitative abstract matching function:
\[
\qlink(\abs{\intent}, \abs{\filter}) =
\begin{cases}
  0 & \link(\abs{\intent}, \abs{\filter}) = 0\\
  1 & \link(\abs{\intent}, \abs{\filter}) = 1\\
  h(\abs{\intent}, \abs{\filter}) & \link(\abs{\intent}, \abs{\filter}) = \top
\end{cases}
\]
where, when static analysis reports a may result, instead of simply
returning $\top$, we return the probability estimated by the model
$h$.

\section{Link-Inference Neural Networks}\label{sec:linn}

\begin{figure}
  \includegraphics[scale=1.2]{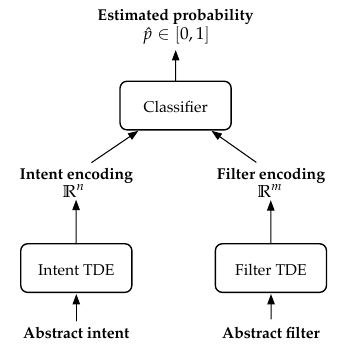}
  \caption{High-level architecture of \linns}
  \label{fig:linn}
  \vspace{-.1in}
\end{figure}

We now present our \emph{link-inference neural network} (\linn)
framework and formalize its components.
Our goal is to take an abstract intent and filter,
$(\abs{\intent},\abs{\filter})$,  representing a may link, and estimate
the probability that the link is a true positive.
\linns are designed to do that.
A \linn is composed of two primary components (see Figure~\ref{fig:linn}):
\begin{itemize}
  \item \textbf{Type-directed encoders} (\tde):
  A \linn contains two different type-directed encoders,
  which are functions mapping an abstract intent or filter
  to a vector of real numbers.
  \emph{Essentially, the encoder acts as a feature
  extractor, distilling and summarizing relevant parts of an intent/filter
  as a real-valued vector.}
  Type-directed encoders are compositions of differentiable functions,
  which can be instantiated using various neural-network architectures.

  \item \textbf{Classification layers}:
  The classification layers take the encoded intent
  and filter and return a probability that there is a link between intent $\intent$
  and filter $\filter$.
  The classification layer we use is a deep
  neural network---\emph{multilayer perceptrons} (\mlp).

  Once we have an encoding of intents and filters
  in $\R^n$, we can use any other classification technique,
  e.g., logistic regression.
  However, we use neural networks because we can
  train the encoders and the classifier simultaneously
  (joint training).
\end{itemize}

Before presenting \tdes in \S\ref{sec:enc}, we \rone present relevant machine-learning background (\S\ref{ssec:ml}) and
 \rtwo describe neural-network architectures
we can use in \tdes (\S\ref{sec:mlp}-\ref{sec:recursive}).
For a detailed exposition of neural networks, refer to~\citet{goodfellow2016deep}.

\newcommand{\rulesep}{\unskip\ \vrule\ }

\subsection{Machine Learning Basics}\label{ssec:ml}
We begin by defining the machine learning problem we aim to
solve and set the notation for the rest of this section.

\paragraph{Machine Learning Problem}
%
Recall that our training data $\train$ is a set
$\{\tpara{(\abs{\intent_i},\abs{\filter_i}),y_i}\}_i,$
where $y_i$ is the label for intent $\abs{\intent}_i$ and filter $\abs{\filter}_i$.
Let $H$ be a \emph{hypothesis space}---a set of possible classifiers.
A loss function $\ell: H \times Z \rightarrow \R$ defines how
\emph{far} the prediction of $H$ is on some item $z \in (\abs{\intents} \times \abs{\filters}) \times Y$.
Our goal is to minimize the loss function over the training
set; we thus define the following quantity: 
$L_\train(h) = \frac{1}{n}\sum_{j=1}^n \ell(h, \tpara{(\abs{\intent_j},\abs{\filter_j}),y_j})$.

Finally, we solve
$\argmin_{h \in H}\ L_\train(h),$
which results in a hypothesis $h \in H$ that minimizes  $L_\train(h)$.
This is typically performed using an optimization
algorithm like \emph{stochastic gradient descent} (\sgd).


\paragraph{Notation}
In what follows,  we  use
$x, y, z, w, \ldots$ to denote column vectors in $\R^n$.
We will use $\vec{x},\vec{y},\vec{z},\vec{w},\ldots$
to denote matrices in $\R^{n\times m}$,
where $\vec{x}_i$ will denote the $i^\emph{th}$ column
vector in matrix $\vec{x}$.
%

\paragraph{Neural Networks}
Neural networks  are transformations described
in the form of matrix operations and pointwise operations.
In general, a neural network $y = g(x;w)$
maps input $x$ to output $y$ with some vector $w$
called the \emph{parameters} of the neural network,
which are to be optimized to minimize the loss function.
There are also other parameterizable aspects of a neural network,
such as the dimensions of input and output,
which need to be determined before training.
Those are called \emph{hyperparameters}.

\subsection{Overview of Neural Architectures}
We now briefly describe a number of neural network
architectures we can use in \tdes.

\paragraph{Multilayer Perceptrons (\mlp)}\label{sec:mlp}
\newcommand{\fc}{\mathit{fc}}
Multilayer perceptrons are the canonical deep neural networks.
They are compositions of multiple \emph{fully connected layers},
which are functions of the form
  $\fc({x}; \vec{w}, {b}) = a(\vec{w}^\top {x} + {b}) : \R^m \to \R^n$,
where $\vec{w}$ and $b$ are parameters,
$m$ is the \emph{input dimension},  $n$ is  the \emph{output dimension}, and $a$ is a pointwise \emph{activation function}.
%
Some common activation functions include \emph{sigmoid} and
\emph{rectified linear unit} (ReLU), respectively:
  $\sigmoid(x) = \frac{1}{1 + e^{-x}}$
and
  $\relu(x) = max(0, x)$.



\paragraph{Convolutional Neural Networks (\cnn)}
\cnns were originally developed for image recognition,
and have recently been successfully applied to natural language processing~\cite{kim2014convolutional, kim2016character}.
\cnns
 are fast to train (compared to \rnns; see below)
and good at capturing \emph{shift invariant} features,
inspiring our usage here for encoding strings.
%
In our setting, we use 1-dimensional \cnns with \emph{global pooling},
transforming a sequence of vectors into a single vector,
i.e., $\R^{n \times l} \to \R^k$, where $l$ is the length of the input and $k$ is the number of \emph{kernels}.
Kernels are functions $\R^{n \times s} \to \R$ of \emph{size} $s$ that are applied to
every contiguous sequence of length $s$ of the input. 
A global pooling function (e.g., $\max$)
then summarizes the outputs of each kernel into a single number.
We use \cnns by
\citet{kim2016character} for encoding natural language,
where kernels have variable size.

\paragraph{Recurrent Neural Networks (\rnn)}\label{sec:recurrent}
%
\rnns
are effective at handling sequential data (e.g., text)
by recursively applying a neural network unit
to each element in the sequence.
An \rnn \emph{unit} is a network
mapping an input and \emph{hidden state} pair at position $i-1$ in the sequence
to the hidden state at position $i$,
that is,
  $\emph{rnn\_unit} : \vec{x}_i,h_{i-1} \mapsto h_i.$
The hidden state $h_i \in \R^n$ can be viewed as a summary of the
sequence seen so far.
%
Given a sequence $\vec{x} \in \R^{m \times l}$ of length $l$,
we can calculate the final hidden state
by applying the \rnn unit at each step.
%
Thus, we view an \rnn as a transformation in $\R^{m \times l} \to \R^n$.

Practically, we use a \emph{long short-term memory network} (\lstm)~\cite{hochreiter1997long},
an \rnn designed to handle long-term dependencies in a sequence.

\paragraph{Recursive Neural Networks}\label{sec:recursive}
Analogous to \rnns, \emph{recursive neural networks}
apply the same unit over recursive structures like trees.
A recursive unit takes input $x_v$ associated with tree node $v$
as well as all the hidden states $\vec{h}_{C(v)} \in \R^{n \times |C(v)|}$ from its children $C(v)$,
and computes the hidden state $h_v$ of the current node.
Intuitively, the vector $h_v$ summarizes the subtree rooted at $v$.

One popular recursive architecture is
\tlstm~\cite{tai2015improved}.
Adapted from \lstms,
a \tlstm unit can
take possibly multiple states from the children of a node in the tree.
For trees with a fixed number $k$ of children per node and/or the order of children matters,
we can have dedicated weights for each child.
This is called a \emph{$k$-ary \tlstm unit}, a function in $\R^{n \times k} \to \R^n$.
For trees with variable number of children per node and/or the order of children does not matter, we can operate over the sum of children states.
This is called a \emph{Child-Sum \tlstm unit}, a function in $\R^{n \times |C(v)|} \to \R^n$ for node $v$.
Refer to \citet[Section 3]{tai2015improved} for details.

%
%


%
\subsection{Type-Directed Encoders}\label{sec:enc}

We are ready to describe how we construct \emph{type-directed encoders} (\tde).
A \tde takes an abstract intent or filter
and turns it into a real-valued vector of size $n$.
As we shall see, however, our construction is generic:
we can take some type $\tau$ and construct
a \tde for $\tau$ by recursively constructing encoders for constituents of $\tau$.
Consider, e.g., a product type comprised of an integer and a string.
An encoder for such type takes a pair of an integer and a string and returns
a vector in $\R^n$.
To construct such an encoder, we compose two separate encoders,
one for integers and one for strings.

\paragraph{Formal Definition}
Formally, an encoder for elements of type $\tau$ is a function
$g : \tau \to \R^n$
mapping values of type $\tau$ into vector space $\R^n$.
For element $a \in \tau$, $g(a)$ is called the encoding of $a$ by $g$.
Integer
$n$ is the dimension of the encoding, denoted by $\dim(g) = n$.

The dimension $n$ of an encoder is
a fixed parameter that we have to choose---a hyperparameter.
The intuition is that the larger $n$ is,
the encoder can potentially capture more information,
but will require more data to train.

\paragraph{Encoder Construction Framework}
In what follows, we begin by describing
encoders for \emph{base types}: reals and characters.
Then, we describe encoders for lists, sets, product types, and sum types.
In our setting of \icc analysis, these types appear
in, for example, the intent action field, which is a list of characters (string) \emph{or} \emph{null},
and filter actions, which are sets of lists of characters (sets of strings).

The inference rules in Figure~\ref{fig:tde}
describe how an encoder $g$ can be constructed for some type
$\tau$, denoted by $g \encs \tau$.
Table~\ref{tbl:enc} lists the parameters used
by a \tde, which are differentiable functions;
the table also lists possible instantiations.
Depending on how we instantiate these functions, we arrive at different
encoders
%
Note that \tdes are compositions of differentiable functions;
this is important since we can jointly train them along with the classifier in \linns.

\begin{example}
Throughout our exposition, we shall use the running
example of encoding an abstract intent $\abs{\intent}$.
We will use $\listt(\tau)$ to denote the type of lists of elements in type $\tau$, and $\sett(\tau)$ to denote sets of elements in $\tau$.
We use $\Omega$ to denote the singleton type that only contains the undefined value $\omega$.

With this formalism, the type of an abstract intent is
$(\listt(\Sigma) + \Omega) \times \sett(\listt(\Sigma))$.
$(\listt(\Sigma) + \Omega)$ is a sum type,
denoting that an action is either a string (list of characters) or \emph{null}.
Type $\sett(\listt(\Sigma))$ is that of categories of an abstract intent---sets of strings.
The product $(\times)$ of action and categories comprises an abstract intent.
Following Figure~\ref{fig:tde},
the construction proceeds in a \emph{bottom-up fashion},
starting with encoders for base types and proceeding upwards.
For instance, to construct an encoder for $\listt(\Sigma)$,
we first require an encoder $g$ for type $\Sigma$,
as defined by the rule $\elist$.
\end{example}

\begin{figure}[t]
    \centering
    \footnotesize

    %

    \prooftree
    \justifies
      \lambda x \ldotp x 
      \encs \R
    \using
        \ereal
    \endprooftree
\hspace{.3in}
%
    \prooftree
      \encchar : \tau_c \rightarrow \R^n
    \justifies
      \encchar 
      \encs \tau_c
    \using
        \echar
    \endprooftree

    \vspace{1em}

    %

    \prooftree
      \begin{array}{c}
      g : \tau \rightarrow \R^n
      \vspace{1pt}
      \\
      \flatten : \listt(\R^n) \rightarrow \R^m
      \vspace{2pt}
    \end{array}
    \justifies
      \lambda x \ldotp \flatten (\map\ g\ x)
      \encs \listt(\tau)
    \using
        \elist
    \endprooftree
    \hspace{1.5em}
%
%
    \prooftree
    \begin{array}{c}
      g : \tau \rightarrow \R^n
      \vspace{1pt}
      \\
      \aggr : \sett(\R^n) \rightarrow \R^m
      \vspace{2pt}
    \end{array}
    \justifies
      \lambda x \ldotp \aggr (\map\ g\ x)
      \encs \sett(\tau)
    \using
        \eset
    \endprooftree

    \vspace{1em}

    \prooftree
      g_1 : \tau_1 \rightarrow \R^n
      \quad
      g_2 : \tau_2 \rightarrow \R^m
      \quad
      \combine : \R^n \times \R^m \rightarrow \R^l
    \justifies
      \lambda (x,y) \ldotp \combine (g_1(x), g_2(y))
       \encs \tau_1 \times \tau_2
    \using
        \eprod
    \endprooftree

    \vspace{1em}

    \prooftree
      g_1 : \tau_1 \rightarrow \R^n
      \quad
      g_2 : \tau_2 \rightarrow \R^n
    \justifies
      \lambda x \ldotp \emph{if}\ \ x \in \tau_1\
      \emph{then}\ \
      g_1(x)\ \ \emph{else}\ \ g_2(x)
       \encs \tau_1 + \tau_2
    \using
        \esum
    \endprooftree

    \vspace{1em}

    $\tau_c$ is a categorical type, e.g., characters.

    \vspace{-1em}

    \caption{Rules for type-directed encoding.
    Representations of functions $\encchar$, $\flatten$, $\aggr$, $\combine$,
    are  in Table~\ref{tbl:enc}.}
    \label{fig:tde}
\end{figure}

\begin{table}[t]
\smaller
  \caption{Types of differentiable functions used in \tdes (Figure~\ref{fig:tde})
  and possible practical instantiations}\label{tbl:enc}
  \vspace{-.1in}
  \begin{tabular}{lll}
    \toprule
        Encoder & Type & Possible differentiable implementations \\
        \midrule
        $\encchar$ & $\Sigma \rightarrow \R^l$ &
        Trainable lookup table (\emph{embedding layer})\\
        $\flatten$ & $\listt(\R^n) \rightarrow \R^m$ & \cnn{} / \lstm \\
        $\aggr$ & $\sett(\R^n) \rightarrow \R^m$ & \emph{sum} / Child-sum \tlstm unit \\
        $\combine$ & $\R^n \times \R^m \rightarrow \R^l$ & Single-layer \mlp{} / binary \tlstm unit\\
        \bottomrule
  \end{tabular}
\end{table}

\subsection*{Encoding Base Types}

We start with encoders for base types.

\paragraph{Numbers}
Since real numbers are already unary vectors,
we can simply use the identity function ($\lambda x \ldotp x$)
to encode elements of $\R$.
This is formalized in the rule $\ereal$.
(The same applies to integers.)

\paragraph{Categorical types (chars)}
We now consider types $\tau_c$ with finitely many values, e.g.,
 \abr{ASCII} characters, Booleans
and user-defined enumerated types.
One common encoding is using a \emph{lookup table}.
Suppose there are $k$ possible values of type $\tau_c$, namely, $a_1, a_2, \dots, a_k$.
The lookup table is often denoted as a matrix $\vec{w} \in \R^{n \times k}$,
where $n$ is the dimension of the resulting encoding.
The encoding for value $a_j$ is simply the $j^{\emph{th}}$ column $\vec{w}^j$ of the matrix $\vec{w}$.

As formalized in rule $\echar$ in Figure~\ref{fig:tde}, we encode
categorical types using an $\encchar$ function,
which we implement as a lookup table whose elements $\vec{w}$
are learned automatically.\footnote{
\emph{One-hot encoding} is a special case of the lookup table, where $W = I^{k \times k}$.}

\begin{example}
  In our example, we need two instances of $\encchar$:
  $\encchar_\Sigma$ for encoding characters in $\Sigma$, and $\encchar_\Omega$ for encoding
  the null type $\Omega$.
\end{example}



\subsection*{\large Encoding Compound Types}

We now switch attention to encoding  compound types.


\paragraph{Lists}
As formalized in rule \elist, to generate an encoder for
$\listt(\tau)$, we require an encoder for type $\tau$.
Given an encoder $g: \tau \rightarrow \R^n$, we can apply
it to every element of the list, thus resulting in a list
of type $\listt(\R^n)$.
Then, using the  function
$\flatten : \listt(\R^n) \rightarrow \R^m$, we can transform
$\listt(\R^n)$ to a vector $\R^m$.
As we indicate in Table~\ref{tbl:enc}, we use
a \cnn or an \lstm to learn the function $\flatten$
during training.

\begin{example}
  To encode strings $\listt(\Sigma)$
  (actions), we construct
  $$\emph{strEnc} = \lambda x \ldotp \flatten (\map\ \encchar_\Sigma\ x)
  \encs \listt(\Sigma)$$
  where $\encchar_\Sigma$ is the encoder we construct for characters,
  and $\map$ is the standard list combinator that applies $\encchar_\Sigma$
  to every element in list $x$, resulting in a new list.
\end{example}

\paragraph{Sets}
%
Rule $\eset$ generates an encoder for type $\sett(\tau)$,
in an analogous manner to $\elist$.
Given an encoder $g: \tau \rightarrow \R^n$, we can apply
it to every element of the set, thus resulting in a set
of type $\sett(\R^n)$.
Then, using the aggregation function
$\aggr : \sett(\R^n) \rightarrow \R^m$, we can transform
$\sett(\R^n)$ to a vector $\R^m$.
As shown in Table~\ref{tbl:enc},
we can simply set $\aggr$ to be the sum of all vectors
in the set,
or use a Child-sum \tlstm unit to pool all elements of a set.

Note that our treatment of sets differs from lists:
We do not use \cnns or \lstms to encode sets, since, unlike lists,
sets have no ordering on elements to  be exploited
by a \cnn or an \lstm.

\begin{example}
  Categories of an intent
  are of type $\sett(\listt(\Sigma))$.
  Given an encoder $\emph{strEnc}$ for $\listt(\Sigma)$,
  we construct the encoder
  $$\emph{catsEnc} = \lambda x \ldotp \emph{aggr} (\emph{map}\ \emph{strEnc}\ x) \encs \sett(\listt(\Sigma))$$
  where $\map$ here applies $\emph{strEnc}$ to every element in  set $x$.
\end{example}

\paragraph{Sum Types}
We now consider sum (union) types $\tau_1 + \tau_2$,
where a variable can take values in $\tau_1$ or $\tau_2$.
Rule $\esum$ generates an encoder for type $\tau_1+\tau_2$;
it assumes that we have two encoders, $g_1$ and $g_2$, for  $\tau_1$ and $\tau_2$, respectively.
Both encoders have to map vectors of the same size $n$.
$\esum$ generates an encoder from
$\tau_1 + \tau_2$
to a vector in $\R^n$.
If a variable $x$ is of type $\tau_1$,
it is encoded as the vector $g_1(x)$.
Alternatively, if $x$ is of type $\tau_2$,
it is encoded as the vector $g_2(x)$.

\begin{example}
  The action field of an abstract intent is a sum type
  $\listt(\Sigma) + \Omega$;
  We construct an encoder for this  as follows:
  $$\emph{actEnc} = \lambda x \ldotp \emph{if}\  x \in \listt(\Sigma)\  \emph{then}\ \emph{strEnc}(x)\ \emph{else}\ \encchar_\Omega(x) \encs \listt(\Sigma) + \Omega$$
\end{example}

\paragraph{Product Types}
We now consider product types of the form $\tau_1 \times \tau_2$.
Rule $\eprod$ generates an encoder for type $\tau_1 \times \tau_2$;
as with $\esum$, it assumes that we have two encoders, $g_1$ and $g_2$, one for type $\tau_1$ and another for $\tau_2$.
Additionally, we assume we have a function
$\combine : \R^n \times \R^m \rightarrow \R^l$.
The encoder for $\tau_1 \times \tau_2$ therefore
encodes type $\tau_1$ using $g_1$, encodes type $\tau_2$
using $g_2$, and then
\emph{summarizes} the two vectors produced by $g_1$ and $g_2$
as a single vector in $\R^l$.
This can be performed using an \mlp, or using a binary \tlstm
unit.
Essentially, we view the the product type as a tree node
with 2 children; so we can summarize the children using a \tlstm unit.

\begin{example}
Finally, to encode an intent, we encode the product type
of an action and categories:
  $$\lambda (a,c) \ldotp \combine(\emph{actEnc}(a), \emph{catsEnc}(b)) \encs
  (\listt(\Sigma) + \Omega) \times \sett(\listt(\Sigma))$$
\end{example}

\def\numTrainingLinks{105,108}
\def\minutesTraining{20}
\def\msPerPredLink{0.23}
\def\FOneScore{0.931}
\def\ROCAUC{0.991}
\def\KrusalsGamma{0.992}

\begin{table}
  \footnotesize
  \centering
  \caption{Instantiations of \tde parameters}
  \label{tab:instantiation}
  \vspace{-.1in}
  \begin{tabular}{llllll}
    \toprule
    \multicolumn{2}{c}{~}     & \multicolumn{4}{c}{\tde parameters}                         \\
      Instantiation                              & Type & $\encchar$ & $\flatten$ & $\aggr$ & $\combine$   \\
    \midrule
    str-\rnn & $\listt(\Sigma)$ & \emph{lookup} & \rnn & - & -  \\
    str-\cnn & $\listt(\Sigma)$ & \emph{lookup} &  \cnn & - & -  \\
    typed-simple & full & \emph{lookup} &  \cnn & \emph{sum} & 1-layer perceptron  \\
    typed-tree & full & \emph{lookup} &  \cnn & \tlstm & \tlstm  \\
    \bottomrule
  \end{tabular}
  {\small
  *\emph{lookup}
  stands for lookup table, as described in \S\ref{sec:enc} and Table~\ref{tbl:enc}.}
\end{table}

\begin{table}
  \footnotesize
  \centering
  \caption{Choices of  hyperparameters}
  \label{tab:hyperparameters}
  \vspace{-.1in}
  \begin{tabular}{lll}
    \toprule
     \multicolumn{2}{c}{Hyperparameter} &  \multicolumn{1}{c}{Choice} \\
    \midrule
    Lookup table & dimension & 16 \\
    \midrule
    \multirow{4}{*}{\cnn} & kernel sizes & $\langle 1, 3, 5, 7\rangle$ \\
    & kernel counts & $\langle 8, 16, 32, 64\rangle$ \\
    & activation & $\relu$ \\
    & pooling & max \\
    \midrule
    \rnn (\lstm) & hidden size  & 128 \\
    \midrule
    \multirow{2}{*}{1-layer perceptron} & dimensions & 64 \\
    & activation & $\relu$ \\
    \midrule
    \multirow{2}{*}{Multilayer perceptron} & dimensions & $\langle 16, 1\rangle$ \\
    & activation & $\langle \relu, \sigma\rangle$ \\
    \bottomrule
  \end{tabular}
\end{table}

\section{Implementation and Evaluation}\label{sec:evaluation}
We now describe an implementation of our technique with
various instantiations and present a thorough evaluation.
We designed our evaluation to answer the
following two questions:
\begin{description}
  \item[Q1: Accuracy] Are \linns
  effective at predicting may links, and what are
  the best instantiations of \tdes for our task?
  \item[Q2: Efficiency] What
   is the runtime performance of model training and link prediction (inference)?
\end{description}
We begin by summarizing our findings:
\begin{enumerate}
  \item The best instantiation of \tdes
  labels may links with an F1 score of
  \FOneScore{}, an area
    under \abr{ROC} curve of \ROCAUC{}, and a Kruskal's $\gamma$  of \KrusalsGamma{}. All these
    metrics indicate high accuracy.
  \item All instantiations complete
  training within $\sim$20min.
  All instantiations take $<$0.2ms
  to label a link,
  except the instantiation using \rnns,
  which takes 2.2ms per link.
\end{enumerate}

\paragraph{Instantiations}
To answer our research questions, we experiment with 4 different instantiations of \tdes,
outlined in Table~\ref{tab:instantiation}, which
are designed to vary in model sophistication,
from simplest to most complex.
In the simplest instantiations, str-\rnn and str-\cnn,  we serialize abstract intents and filters as strings ($\listt(\Sigma)$)
and use \rnns and \cnns for encoding.
In the more complex instantiations, typed-simple and typed-tree,
we maintain the \emph{full} constituent types for intents and filters, as illustrated in \S\ref{sec:enc}.
The typed-simple instantiatition uses the simplest choices for
$\aggr$ and $\combine$---sum of all elements and a single-layer neural network, respectively.
The typed-tree instantiations uses Child-Sum and binary \tlstm units instead.

\paragraph{Hyperparameters}
Table~\ref{tab:hyperparameters} summarizes the choice of \emph{hyperparameters}.
We choose the embedding dimension and the range of kernel sizes
following a similar choice as in \citet{kim2016character}.
We choose only odd sizes for alignment consideration.
We choose $\relu$ for all activations as it is both effective in prediction and efficient in training.
For global pooling in \cnns, $\max$ is used as
we are only interested in the existence of a pattern,
rather than its frequencies.
The hidden size of \rnns is set close to the output size of \cnns,
so we could fairly compare the capability of the two architectures.
\tlstms used in our instantiations require no hyperparameterization,
 as the hidden size is fixed to the output size of the string encoder.

\paragraph{Implementation}
We have implemented our instantiations in Python using Keras~\cite{chollet2015} with the TensorFlow~\cite{abadi2016tensorflow}
backend.
We chose \emph{cross entropy} as our loss function
and  RMSprop \cite{tieleman2012lecture}, a variation of stochastic gradient descent (\sgd), for optimization.
All experiments were performed on a machine with an Intel Core
i7-6700 (3.4 GHz) CPU, 32GB memory, 1TB SSD, and Nvidia GeForce GTX 970 GPU.
The \linn was trained and tested on the GPU.

\subsection{Experimental Setup}
\paragraph{Corpus}
We used the \abr{PRIMO}~\cite{octeau16}  corpus,
 consisting of 10,500 Android applications from Google Play. For static analysis, we
used \icthree~\cite{old+15} combined
with \abr{PRIMO}'s abstract matching procedure.
This combination
provides a definitive \emph{must}/\emph{may} label to pairs of
 abstract intents and
filters.


\paragraph{Simulating Imprecision}
The challenge with setting up the training and test
data is that we do not know the \emph{ground truth}
label of may links.
To work around that, we adopt the ingenious approach
of \citet{octeau16}, who construct synthetic may links from must
links by instilling imprecisions in abstract intents and filters.
This results in a set of may links for which we know the label.
Formally, to add imprecision to an abstract intent $\abs{\intent}$,
we transform it into a \emph{weaker} abstract value $\abs{\intent_w}$
such that $\abs{\intent} \subseteq \abs{\intent_w}$.
Intuitively,  adding imprecisions involves
adding regular expressions, like $\texttt{".*"}$
into, e.g., action strings.

Following~\cite{octeau16},
we use the empirical distribution of imprecision observed
in the original abstract intents (filters) to guide the imprecision simulation process.
Specifically, we distinguish the different types of imprecision (full or
partial, and by each field) and calculate the empirical
probability for each of them as they appear in our data.
These probabilities are then used to introduce
respective imprecisions in the intents and filters.  Note that the introduction of
imprecision does not necessarily convert a must link into a may link---since even if we weaken an abstract intent,
the abstract matching process may still detect its label
definitively.

\paragraph{Sampling and Balancing}
We train the model on a sampled subset of links, as the number of links is
quadratic in the number of intents and filters
and training is costly (we have 3,856 intents and 37,217
filters, leading to over 100 million links). Sampling
may result in loss of information; however, as we show,
our models perform well.

Another important aspect of sampling is balancing.
Neural networks 
are sensitive to a balanced presence of positive and negative training instances.
In our setting, links over  some particular intent (filter)
could easily fall into one of two extreme cases where
most of them are positive (intent is vague)
or most of them are negative (intent is specific).
So we carefully sample links balanced
between positive and negative labels,
and among intents (filters).
Sampling and balancing led to
a training set consisting of 105,108 links,
(63,168 negative and 41,940 positive),
and a testing set consisting of 43,680 links
(29,260 negative and 14,420 positive).
\emph{Note that the testing set is comprised
solely of may links}.

\begin{figure}
  \small
  \centering
  \begin{subfigure}[b]{0.22\textwidth}
    \centering
    \includegraphics[scale=0.4]{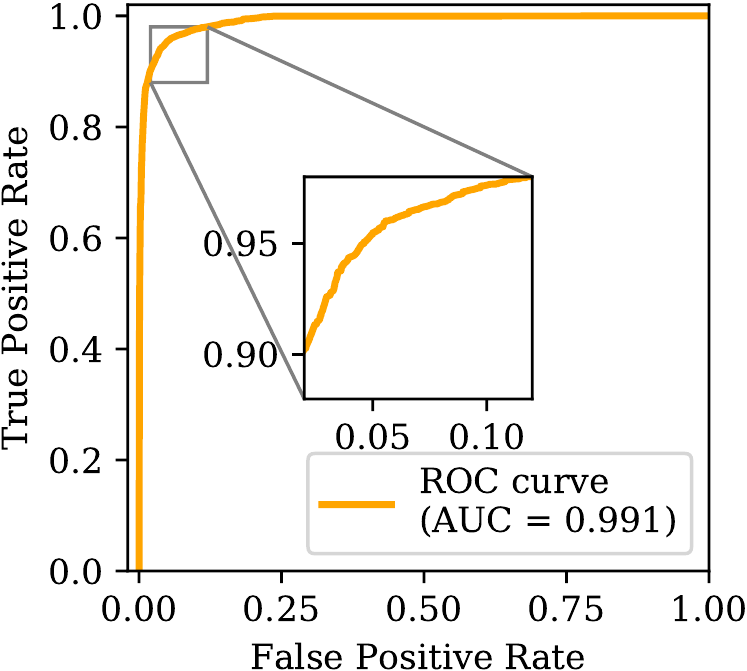}
    \caption{Receiver operating characteristic (\abr{ROC})}
    \label{fig:roc-curve}
  \end{subfigure}
  \hspace{0.02\textwidth}
  \begin{subfigure}[b]{0.22\textwidth}
    \centering
    \includegraphics[scale=0.4]{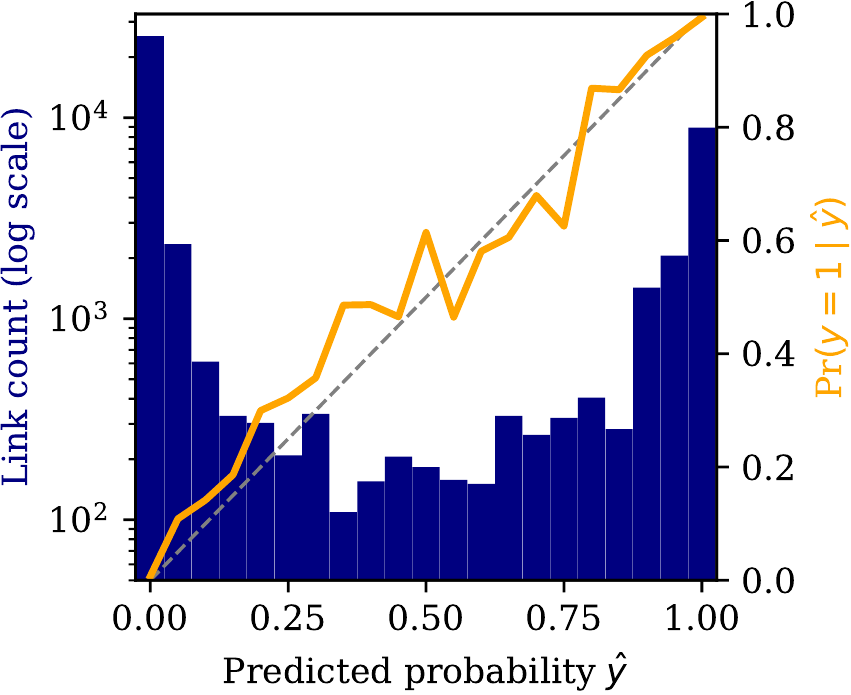}
    \caption{Distribution of predicted link probabilities}
    \label{fig:prediction-distribution}
  \end{subfigure}
  \vspace{-0.1in}
  \caption{Detailed results for the typed-tree instantiation}
  \vspace{-0.1in}
  \label{fig:results}
\end{figure}

\begin{table*}
  \footnotesize
  \centering
  \caption{Summary of model evaluations}
  \label{tab:results-summary}
  \vspace{-.1in}
  \begin{tabular}{l|cccccccc}
    \toprule
    Instantiation & \# Parameters & Inference time ($\mu$s/link) & Testing $\gamma$ & Testing F1 & \abr{AUC} & Entropy of $\hat{y}$ & $\emph{Pr}(y=1 \mid \hat{y} > 0.95)$ & $\emph{Pr}(\hat{y} > 0.95)$ \\
    \midrule
    str-\rnn   & 154,657 & 2220 & 0.970 & 0.891 & 0.975 & 3.002 & 0.980 & 0.089 \\
    str-\cnn   & \textbf{27,409}  & \textbf{57}   & 0.988 & 0.917 & 0.988 & 2.534 & \textbf{0.998} & 0.139 \\
    typed-simple    & 142,417 & 157  & 0.989 & 0.920 & 0.988 & 2.399 & 0.996 & 0.173 \\
    typed-tree & 634,881 & 171  & \textbf{0.992} & \textbf{0.931} & \textbf{0.991} & \textbf{2.220} & 0.994 & \textbf{0.200} \\
    \bottomrule
  \end{tabular}
\end{table*}

\subsection{Results}

\paragraph{Q1: Accuracy}
To measure the predictive power of our
four instantiations, we use a number standard metrics:
\begin{itemize}
  \item \emph{F1 score} is the harmonic mean of \emph{precision} and \emph{recall},
  which gives a balanced measure of a predictor's false-negative and false-positive rates.
  When computing F1, the model output is rounded to get a 0/1 label.
  A value of 1 indicates perfect precision/recall.

  \item \emph{Receiver operating characteristic (\abr{ROC}) curve}
  plots the true-positive rate
  against the false-positive rate.
  It represents a binary classifier's diagnostic performance
  at different discrimination thresholds.
  A perfect model has an area under the curve (\abr{AUC})
  of 1.

  \item \emph{Kruskal's $\gamma$} is used to measure the correlation between the ranking computed by our model and the ground truth on may links, which is useful in our setting as
in practice we would use computed probabilities to
rank may links in order of likelihood to present them to the user for investigation.
   A value of 1 indicates perfect correlation between
   computed ranking and ground truth.
\end{itemize}

The metrics for each instantiation are summarized in Table~\ref{tab:results-summary}.
In addition to the aforementioned metrics,
we  include the number of trainable parameters, indicating the fitting power (lower is better);
inference time (time to evaluate one instance);
entropy of predicted probabilities $\hat{y}$, indicating the power of triaging links (lower is better);
probability of true positives within links with high predicted values $\emph{Pr}(y=1 \mid \hat{y} > 0.95)$ (higher is better);
and portion of links with such high predicted values $\emph{Pr}(\hat{y} > 0.95)$ (higher is better).

We observe that the best instantiation is typed-tree.
Note, however, that using \tlstm involves the largest
number of trainable parameters.
With the fewest trainable parameters and fastest running time,
the str-\cnn model achieves the highest true-positive rate
within the most highly ranked links.
The str-\rnn model is worse than others in almost all aspects,
suggesting that the \rnn string encoder struggles to capture useful patterns from intents (filters).
Our results show that more complex models, specifically those that use more
parameters or encode more
structure, tend to perform slightly better.
\emph{While simple models are
good enough,  more complex models may still lend significant advantage when we
consider market-scale analysis: link inference has to run on millions of
links and even small differences (e.g. a false positive rate difference of just
0.49\% between our two best models, typed-tree and typed-simple) would translate to thousands of mislabeled links.}
%
%

%
Figure~\ref{fig:roc-curve} shows the \abr{ROC} curve over the testing dataset for the typed-tree model.
Figure~\ref{fig:prediction-distribution} shows the distribution of the predicted probabilities of the existence of a link,
and the  probability of a link actually existing
given the predicted probability.
The curve depicting the conditional probability would ideally follow $y=x$ line.
Therefore, the model's prediction value is highly correlated to the true probability of being a positive,
which confirms with the high $\gamma$ value we observed.
A higher value at the upper right of the conditional probability curve suggests
links highly ranked ($\hat{y}$ > 0.95) by our models are highly likely actual positives.
This is particularly important in saving the effort of humans involved in investigating links.


Compared to \abr{PRIMO}~\cite{octeau16}, our evaluation is over a set of may links only.  This is a
strictly more difficult setting than \abr{PRIMO}'s, which evaluated over all links containing any
imprecision---so some of the links are actually must links. In particular, if a link is \emph{must not exist}, a partially
imprecise version will also mostly be \emph{must not exist}. For example,
\code{a(.*)c} does not match \code{xyz}.
\emph{Despite working in a more challenging setting,
our best instantiations are still able to achieve a Kruskal's $\gamma$ value $(0.988-0.992)$ higher than \abr{PRIMO} $(0.971)$}
(In fact, if we use the same setting as \abr{PRIMO}, our Kruskal's $\gamma$
reaches $0.999$.)

\paragraph{Q2: Efficiency}
Regarding running time,
all our instantiations, except str-\rnn, are efficient.
As shown in Table~\ref{tab:results-summary},
they take no more than 171\textmu s per link for inference.
The training time is proportional to inference time.
The three best instantiations take no more than 20min
to finish 10 \emph{epochs} of training.
We consider this particularly fast, given that we used only average
hardware and deep-neural-network training often lasts several hours or days for
many problems.
The only exception is str-\rnn,
which is about 13 times slower than the slowest of the other three,
mainly because of the inefficient unrolling of \rnn units over long input strings.

The storage costs or the model size
can be measured in the number of trainable parameters.
As shown in Table~\ref{tab:results-summary},
the largest model (typed-tree) consists of about 634K parameters,
taking up 5.6 MB on disk;
the smallest model (string-\cnn) consists of only about 27K parameters,
taking up merely 300KB on disk. In summary, even the most complex model does
not have significant storage cost.


%
\paragraph{Threats to Validity}\label{sec:discussion}
We see two threats to validity of our evaluation.
First, to generate labeled may links for testing,
we synthetically instilled imprecision, following
the methodology espoused by \abr{PRIMO}, enabling a head-to-head comparison with
\abr{PRIMO}.
While the synthetic imprecision follows the empirical imprecision
distribution, we could imagine
that  actual may links only exhibit rare, pathological imprecisions.
While we cannot discount this possibility, we believe it is unlikely.
The ultimate test of our approach is in improving the efficacy of client
analyses such as IccTA~\cite{lbb+15} and we intend to investigate the impact of
client analyses in future work.

Second, with neural networks, it is often unclear
whether they are capturing relevant features
or simply \emph{getting lucky}. To address this
threat, we next perform an interpretability investigation.

\section{Interpretability Investigation}\label{sec:explainability}
We now investigate whether our models are learning the right things and not
basing predictions on extraneous artifacts in the data. An informal way to frame
this concern is \emph{whether the model's predictions align with human
intuition}. This
is directly associated with the trust we can place on the model and its
predictions~\cite{rsg16}.

Our efforts towards interpretability are multifold:

\begin{itemize}
  \item  We analyzed the incorrect classifications to \emph{understand the reasons
    for incorrectness}. Our analysis indicates that the reasons for incorrectness
    are understandable and intuitively explained.
    (\S\ref{sec:error-inspection})
  \item We analyze several instances
    to \emph{understand what parts of input are important for the predictions}. Results
    show that the model is taking expected parts of the inputs into account.
    (\S\ref{sec:explaining-instances})
  \item We \emph{studied activations of various kernels inside the neural
    networks}
    to see which inputs activated them the most. Our findings show several
    recognizable strings, which intuitively should have a high influence on
    classification, to be activating the kernels.
    (\S\ref{sec:kernel-activations})
  \item  We used t-SNE visualizations~\cite{maaten2008visualizing}
    to \emph{understand how a model groups similar inputs and what features it
    may be using}. We found several important patterns,
    boosting our trust in the model.
    (\S\ref{sec:encodings})
\end{itemize}

Interpreting and explaining predictions of a ML-algorithm is a challenging
problem and in fact a topic of active research~\cite{xai,xai2}. Our study in
this section is thus a \emph{best-effort} analysis.

\subsection{Error Inspection}\label{sec:error-inspection}
We present anecdotal insights from our investigation of erroneous classification
in the typed-simple model.
A number of misclassifications appear where the action or category fields are
completely imprecise (e.g., action string is \code{(.*)}) or too
imprecise to make meaningful deductions. The model unsurprisingly has difficulty
in classifying such cases.

There are also, albeit fewer, misclassified instances
for near-certain matches. Examples include
\longcode{android.intent.ac\-tion.VIEW} matching against
\longcode{android.intent.action(.*)EW}  and
\longcode{android.in\-tent.action.GET\_C(.*)NTENT} matching against
\longcode{android.intent. action.GET\_CONTENT} (as part of intents and filters).
Such misses may be due to imprecision preventing  encoders
from detecting characteristic patterns (e.g. \code{(.*)EW} is seemingly very
different from \code{VIEW}).
%
%
Finally, Our model predicts a matches
for different but very similar strings. For example, consider
\longcode{com.andromo. dev48329.app53751.intent.action.FEED\_STAR(.*)ING} and
\longcode{com. andromo.d(.*)9.app135352.intent.action.FEED\_STARTING}. The two look
similar but have different digits
We explore this case more in \S~\ref{sec:encodings}.

Overall, our model is able to extract and generalize useful patterns.
However, it probably has less grasp on the exact meaning of
\code{(.*)}, so there are failures to identify some highly probable matches.

\begin{figure}
  \includegraphics[width=0.97\columnwidth]{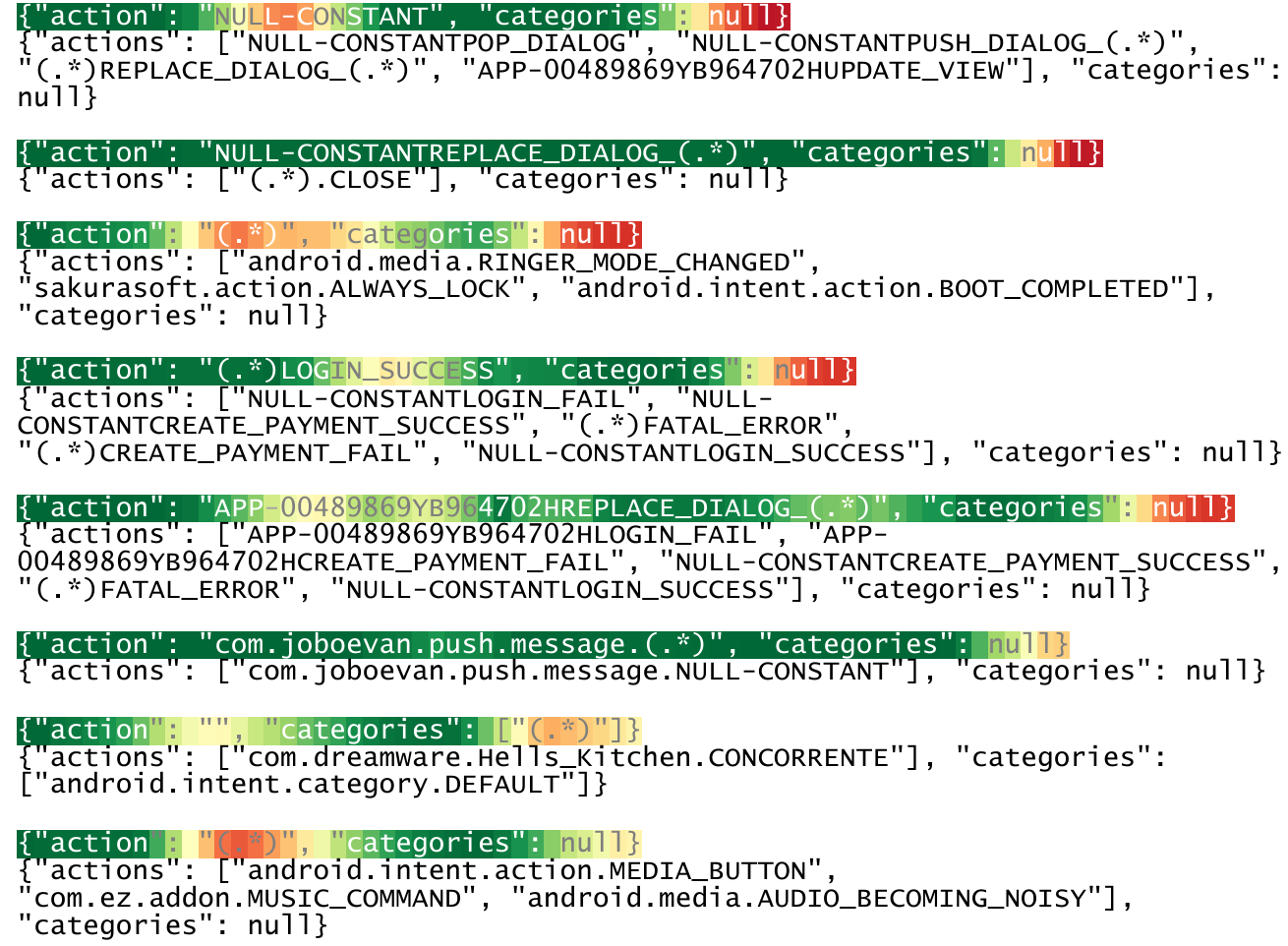}
  \vspace{-.13in}
  \caption{Explaining individual instances}
  \vspace{-.1in}
  \label{fig:explainability}
\end{figure}

\subsection{Explaining Individual Instances}\label{sec:explaining-instances}
We would like to explain predictions on
individual instances. Our approach is similar to that of the
popular system \abr{LIME}~\cite{rsg16}: we
perturb the given instance and study change in prediction of the model, treated as a
black box, due to the perturbation. In this way, we
can ``explain'' the prediction on the given instance. We used the str-\cnn
instantiation and perturbed the string representation of the given intent while keeping
the filter string constant. Our perturbation technique is to mask (delete) a substring
from the string. We perform this perturbation at all
locations of the string. Using a mask length of 5, the examples in
Figure~\ref{fig:explainability} show intents and filters with a heat map
overlaid on the intents while
keeping the filters uncolored. All the instances selected here have positive
link predictions (it is unlikely to perturb a negatively predicted link to
result in a positive prediction). The redder regions indicate a
significant difference in prediction when masking around those
regions, thus indicating that the regions are important for the predictions.

The examples generally show red regions in the action and categories values,
which implies that those values are important for the prediction. Thus, the
model is generally making good decisions about what is important. There are a
couple of outliers. First, the second instance does not have the action field
reddened. This is probably because there is no direct match between the action
strings in the intent and filter and it may just be the case that the model was
learnt from similar examples in the training set. The other outlier is the
third-last example: we believe that masking any small part of the action still
leaves enough useful information, which the model can pick up to make a positive
prediction. Overall, these examples give us confidence that the model has
acquired a \emph{reasonably correct} understanding of intent resolution.

\subsection{Studying Kernel Activations}\label{sec:kernel-activations}
We tested input strings on the str-\cnn instantiation to find
if the \cnn kernels are picking up patterns
relevant to \icc link inference.
We went over all intent strings in our dataset
and for each \cnn kernel, looked for string segments that activate the kernel the most.
Some representative kernels and their top activating segments
are shown in Table~\ref{tab:top-kernel-activations}.
Important segments are noticeable.
For example, kernel \code{conv1d\_size5:14} gets activated on \code{(.*)},
and kernel \code{conv1d\_size7:0} gets activated on \code{VIEW}.
Kernel \code{conv1d\_size5:3} is interesting,
as it seems to capture \code{null},
an important special value, but also captures \code{sulle}.
This indicates a single kernel on its own may not be so careful at distinguishing
useful vs. less useful but similar looking segments.
Apart from such easily interpretable cases,
many other activations appear for seemingly meaningless strings.
It is likely that in combination, they help capture subtle context,
and when combined in later layers, generate useful encodings.

\begin{table}
  \footnotesize
  \centering
  \caption{Some \cnn kernels and their top stimuli}
  \label{tab:top-kernel-activations}
  \vspace{-.1in}
  \begin{tabular}{cc|cc|cc}
    \toprule
    \multicolumn{2}{c}{\code{conv1d\_size5:14}} & \multicolumn{2}{c}{\code{conv1d\_size5:3}} & \multicolumn{2}{c}{\code{conv1d\_size7:0}} \\
    segment & activation & segment & activation & segment & activation \\
    \midrule
    \code{(.*)R} & 1.951 & \code{null\}} & 3.796 & \code{TAVIEWA} & 3.704 \\
    \code{(.*)u} & 1.894 & \code{null,} & 2.822 & \code{n.VIEW"} & 3.543 \\
    \code{(.*)t} & 1.893 & \code{sulle} & 2.488 & \code{y.VIEW"} & 3.384 \\
    \bottomrule
  \end{tabular}
  \vspace{-.1in}
\end{table}

\subsection{Visualization of Encodings}\label{sec:encodings}
We present a visualization to discover interesting patterns in encodings.
In the
interest of space, we only discuss intent encodings for the typed-simple instantiation.
We use t-SNE~\cite{maaten2008visualizing}, a non-linear dimensionality reduction
technique frequently used for visualizing high-dimensional data in two or three
dimensions. The key aspect of the technique is that similar objects
are mapped to nearby points while
dissimilar objects are mapped to distant points.
Figure~\ref{fig:intent-encoding} shows the encodings
for all intents in the training set (sub-figures highlight different areas).
The size of each point reflects the number of intents
sharing the same combination of values.

\def\cloudHighlightedSubfigure{0.47\textwidth}
\def\cloudHighlightedScale{0.27}
\begin{figure}[t]
  \centering
  \begin{subfigure}[b]{0.4\textwidth}
    \centering
    \begin{subfigure}[t]{\cloudHighlightedSubfigure}
      \centering
      \includegraphics[scale=\cloudHighlightedScale]{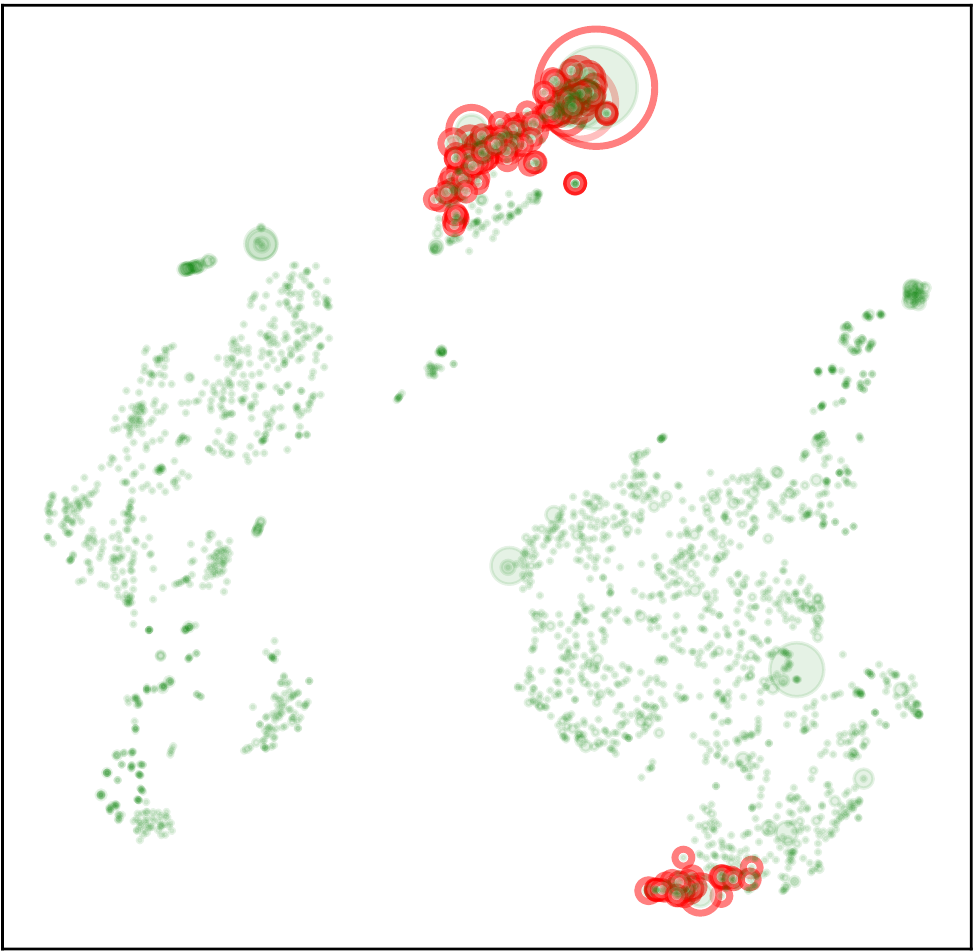}
      \caption{\code{android.intent.*}}
      \label{fig:intent-encoding-highlighted-by-android-intent}
    \end{subfigure}
    \hfill
    \begin{subfigure}[t]{\cloudHighlightedSubfigure}
      \centering
      \includegraphics[scale=\cloudHighlightedScale]{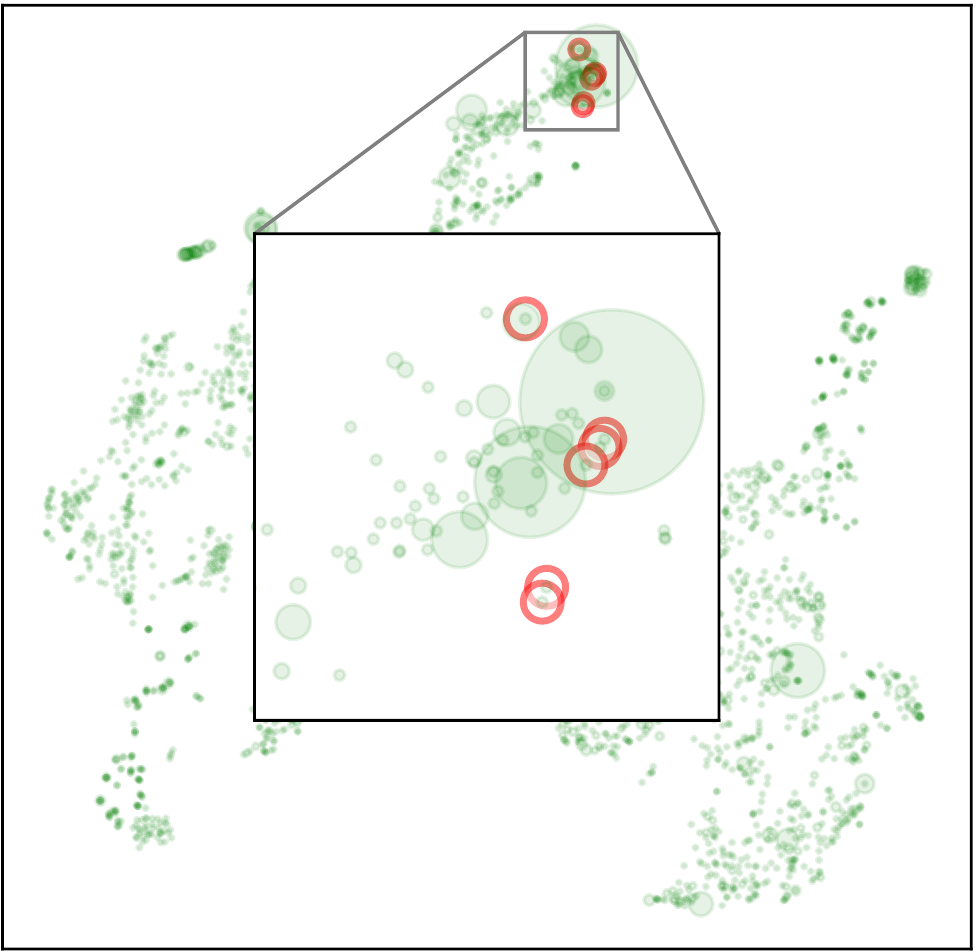}
      \caption{Imprecise \code{VIEW} actions}
      \label{fig:intent-encoding-highlighted-by-view}
    \end{subfigure}
    \\\vspace{0.1in}
    \begin{subfigure}[t]{\cloudHighlightedSubfigure}
      \centering
      \includegraphics[scale=\cloudHighlightedScale]{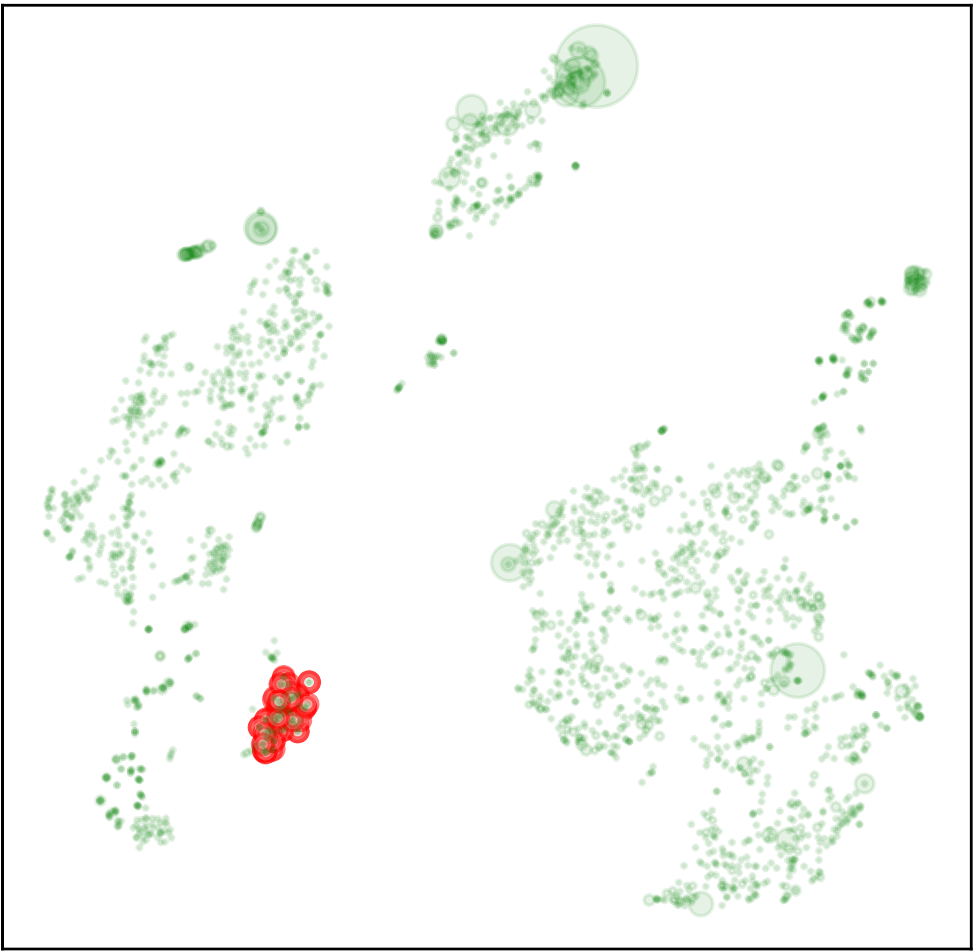}
      \caption{\code{dev*.app*.*.FEED*}}
      \label{fig:intent-encoding-highlighted-by-andromo-feed}
    \end{subfigure}
    \hfill
    \begin{subfigure}[t]{\cloudHighlightedSubfigure}
      \centering
      \includegraphics[scale=\cloudHighlightedScale]{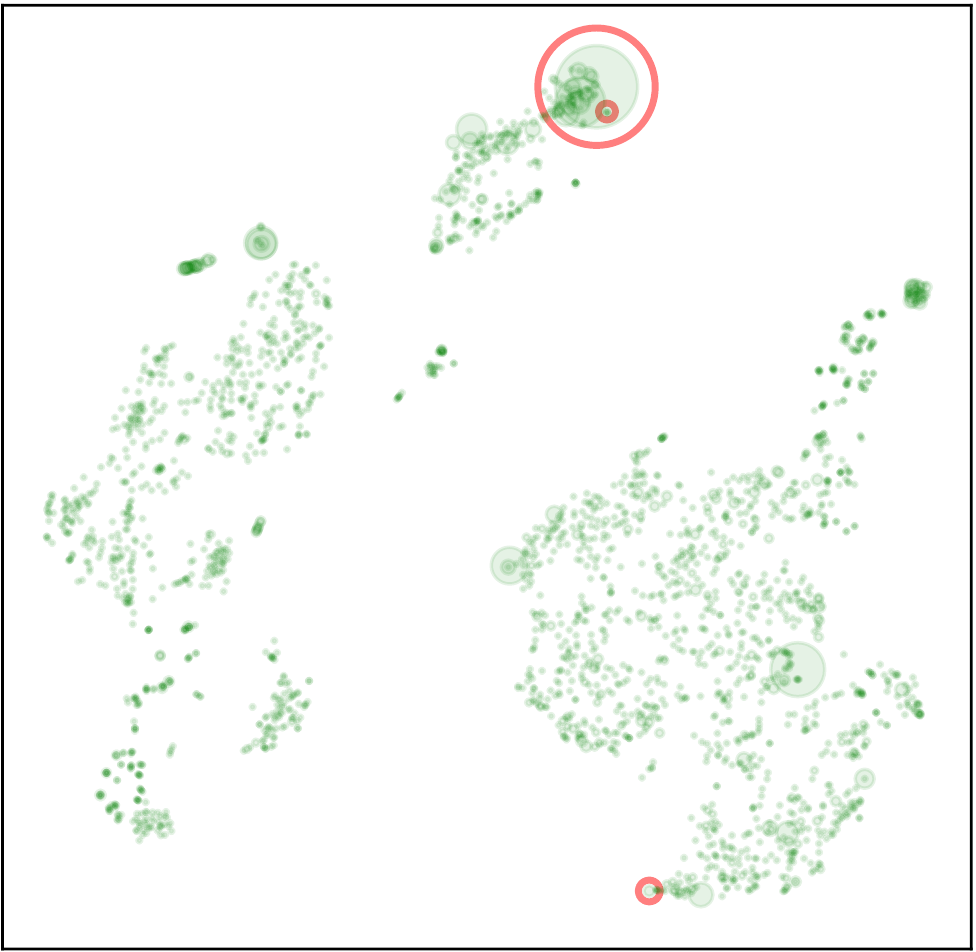}
      \caption{\code{DEFAULT}, total imprecise and null categories}
      \label{fig:intent-encoding-highlighted-by-categories}
    \end{subfigure}
  \end{subfigure}
  \vspace{-0.1in}
  \caption{Intent encodings visualized using t-SNE}
  \label{fig:intent-encoding}
  \vspace{-0.1in}
\end{figure}

Figure~\ref{fig:intent-encoding-highlighted-by-android-intent}
highlights the intents for which the action starts with \longcode{android.intent}.
The top cluster is associated with category \longcode{android.intent.category.DEFAULT}
and the bottom is associated with \code{null} categories.
A few different actions comprise the top cluster, e.g.,
\longcode{android.intent.action.VIEW}, \longcode{android.intent.action.SEND}, and
imprecise versions of these actions.
Figure~\ref{fig:intent-encoding-highlighted-by-view} zooms in on the top
cluster to show six imprecise versions of the \code{VIEW} action:
one top circle
containing \texttt{android.inte(.*).VIEW};
three  circles containing
\texttt{android.intent(.*)n.VIEW}, \texttt{android.intent.(.*)n.VIEW}, and
\texttt{android.intent(.*).VIEW}; and two
lower-figure circles containing
\texttt{an\-(.*)ction.VIEW} and \texttt{a(.*)action.VIEW}.  We can thus see the level
of imprecision reflected spatially,
while the semantic commonality is still well preserved
as the points are all close to the big circle of the precise value.

Figure~\ref{fig:intent-encoding-highlighted-by-andromo-feed} highlights actions
whose last part starts with \code{FEED}.
The prefixes are largely different
but share the pattern of \code{dev*.app*},
where \code{*} are strings of digits.
Our model is thus able to extract the common pattern and
generate similar encodings for these intents.

The last example is the effect of three important values of the categories field:
\{\longcode{android.intent.category.DEFAULT}\},
total imprecision \{\code{(.*)}\},
and \{ \}.
In
Figure~\ref{fig:intent-encoding-highlighted-by-categories},
we highlight three encodings with the \code{VIEW} action
and the above categories.
The encoding for that of \{ \} (bottom circle)
is separated from the other two (top circle).
This conforms with the fact that
total imprecision is the imprecise version of some concrete value,
which is most likely \code{DEFAULT},
rather than the absence of the value for the categories field,
which means ignoring the field during resolution.
Our observations suggest that our intent encoder is able
to automatically learn semantic (as opposed to merely syntactic) artifacts.


%

\section{Related Work}\label{sec:relatedwork}

\paragraph{Android Analysis}
The early ComDroid system~\cite{cfgw11}
inferred subsets of \icc values using simple, {\it ad hoc}
static analysis. Epicc~\cite{omj+13}
constructed a more detailed model of \icc.
\icthree~\cite{oljp16} and \citet{blyw17} improved on Epicc by using better
constant propagation.
These static-analysis techniques inevitably overapproximate the set of possible
\icc values and can thus potentially benefit from our work.
\abr{PRIMO}~\cite{octeau16} was the first to
formalize the \icc link resolution process. We have compared
with \abr{PRIMO} in Sections~\ref{sec:introduction} and \ref{sec:evaluation}.

\icc security studies range from information-flow
analysis~\cite{blyw17,gkp+15,kfb+14,lbb+15,wror14} to vulnerability
detection~\cite{bsgm15,li2014automatically,llw+12,fada14}.
These works do not perform a formal \icc analysis and are instead consumers of \icc analysis and can potentially benefit from our work.

\paragraph{Static Analysis Alarms}
Z-Ranking~\cite{ke03} uses a simple counting technique to
go through analysis alarms and rank them.
Since bugs are rare,
if the analysis results are mostly safe, but then
an error is reported, then it is likely an error.
\citet{tgpa14} train a classifier using manually supplied
labels and feature extraction of static analysis results.
We utilize static analysis to automatically do the labeling, and perform feature extraction automatically from abstract values.
\citet{koc2017learning}
use manually labeled data and neural networks
to learn code patterns where static analysis
loses precision. Our work uses automatically generated
must labels, and since errors (links) are not localized in one line or even application,
we cannot pinpoint specific code locations for training.
Merlin~\cite{Livshits:2009:MSI:1542476.1542485}
infers probabilistic information-flow specifications
from code and then applies them to discover bugs;
similar approaches were also proposed by
\citet{Kremenek:2006:UBI:1298455.1298471} and Murali et
al.~\cite{murali2017bayesian}.
We instead augment static analysis with a post-processing
step to assign probabilities to alarms.

Another class of methods uses logical techniques
to suppress false alarms, e.g., using abduction~\cite{Dillig:2012:AED:2254064.2254087}
or predicate abstraction \cite{Blackshear:2013:ASM:2491956.2462188}.

\paragraph{ML for Programs}
There is a plethora of works applying forms of machine learning
to ``big code''---see \citet{allamanis2017survey}
for a comprehensive survey. We compare with most related works.

Related to \tdes, \citet{parisotto2016neuro} introduced
a tree-structured neural network to encode abstract-syntax trees
for program synthesis.
\tdes can potentially be
extended in that direction to reason about recursive data types.
\citet{pmlr-v48-allamanis16} use
\cnns to generate summaries of code.
Here, we used \cnns to encode list data types.

\citet{allamanis2017learning}
present neural networks
that encode and check semantic equivalence of two symbolic formulas.
The task is more strict than link inference. 
For our case, simple instantiations work reasonably well. 
It is also prohibitive to generate all possible intents or filters
as they did for Boolean formulas. 

Hellendoorn and Devanbu~\cite{hellendoorn2017deep} proposed modeling techniques
for source code understanding, which they test with n-grams and \rnns. Gu
et al.~\cite{gu2016deep} use deep learning to generate \abr{API} sequences for a
natural language query (e.g., parse \abr{XML}, or play audio). Instead of using deep
learning for code understanding, we use it to classify imprecise results of static
analysis for inter component communication.
\citet{raychev2014code} use \rnns for code completion
and other techniques for predicting variable names and types~\cite{Raychev:2015:PPP:2676726.2677009}.

\section{Conclusions}\label{sec:conclusions}

We tackled the problem of false positives in static analysis of Android communication links. We
 augment a static analysis with a post-processing step that
estimates the probability with which a link is indeed a true positive.
 To facilitate machine learning
in our setting, we propose a custom neural-network architecture
targeting the link-inference problem.  The key novelty
is \emph{type-directed encoders} (\tde),
a framework for composing neural networks that take elements of some type
$\tau$ and encodes them into real-valued vectors.  We believe that
this technique can be applicable to a wide variety of problems in the
context of machine-learning applied to programming-languages problems. In the future, we want to explore this intriguing idea in other contexts.

\begin{acks}
  This material is partially supported by the National Science Foundation
  Grants CNS-1563831 and CCF-1652140 and by the Google Faculty Research Award. Any opinions,
  findings, conclusions, and recommendations expressed herein are those
  of the authors and do not necessarily reflect the views of the funding
  agencies.
\end{acks}

\balance
\bibliography{bibfile}

%

\end{document}